\documentclass[journal]{IEEEtran}
\usepackage{cite}
\usepackage{graphicx}
\usepackage{amsmath}
\usepackage{amssymb}
\usepackage{booktabs}
\usepackage{bbm}
\usepackage{color}
\ifCLASSINFOpdf
\else
\fi

\begin{document}
\title{Learning Music-Dance Representations through Explicit-Implicit Rhythm Synchronization}
% ~\IEEEmembership{Member,~IEEE,}
\author{Jiashuo Yu, Junfu Pu*, Ying Cheng, Rui Feng*, and Ying Shan% <-this % stops a space
\thanks{Jiashuo Yu and Rui Feng are with the School of Computer Science, Shanghai Key Lab of Intelligent Information Processing,
Shanghai Collaborative Innovation Center of Intelligent Visual Computing, Fudan University, 
Shanghai, 200438, China (E-mail: \{jsyu19, fengrui\}@fudan.edu.cn.)}% <-this % stops a space
\thanks{Ying Cheng is with the Academy for Engineering and Technology, Fudan University, Shanghai, 200438, China (E-mail: chengy18@fudan.edu.cn.)}
\thanks{Junfu Pu and Ying Shan are with the Applied Research Center, PCG, Tencent, Shenzhen,
 518000, China (E-mail: \{jevinpu, yingsshan\}@tencent.com.)}% <-this % stops a space
\thanks{This work was done when Jiashuo Yu was an intern in ARC Lab, Tencent PCG.}% <-this % stops a space
\thanks{*Corresponding authors.}}


\markboth{IEEE TRANSACTIONS ON MULTIMEDIA, VOL. **, NO. *, JAN 2023}%
{}



\maketitle

\begin{abstract}
   Although audio-visual representation has been proven to be applicable in many downstream tasks, the representation of dancing videos, which is more specific and always accompanied by music with complex auditory contents, remains challenging and uninvestigated. Considering the intrinsic alignment between the cadent movement of the dancer and music rhythm, we introduce \textbf{MuDaR}, a novel \textbf{Mu}sic-\textbf{Da}nce \textbf{R}epresentation learning framework to perform the synchronization of music and dance rhythms both in explicit and implicit ways. Specifically, we derive the dance rhythms based on visual appearance and motion cues inspired by the music rhythm analysis. Then the visual rhythms are temporally aligned with the music counterparts, which are extracted by the amplitude of sound intensity. Meanwhile, we exploit the implicit coherence of rhythms implied in audio and visual streams by contrastive learning. The model learns the joint embedding by predicting the temporal consistency between audio-visual pairs. The music-dance representation, together with the capability of detecting audio and visual rhythms, can further be applied to three downstream tasks: (a) dance classification, (b) music-dance retrieval, and (c) music-dance retargeting. Extensive experiments demonstrate that our proposed framework outperforms other self-supervised methods by a large margin.
\end{abstract}


% \fi
\IEEEpeerreviewmaketitle

\section{Introduction}
\label{sec:intro}

Recent years have witnessed the rapid growth of dancing video amounts on online video-sharing websites. The demand for automatically processing dancing videos based on content is thence becoming increasingly stronger. In literature, many prior works have achieved promising results on music-dance tasks, e.g., music-driven dance generation~\cite{zhuang2020music2dance, lee2019dancing, guo2021danceit, huang2020dance} and music-dance alignment~\cite{wang2020alignnet}. However, these methods are usually task-specific and have a strong dependence on labeled data, which restricts their applicability to a large extent. 

\begin{figure}[t]
\centering
\includegraphics[width=0.48\textwidth]{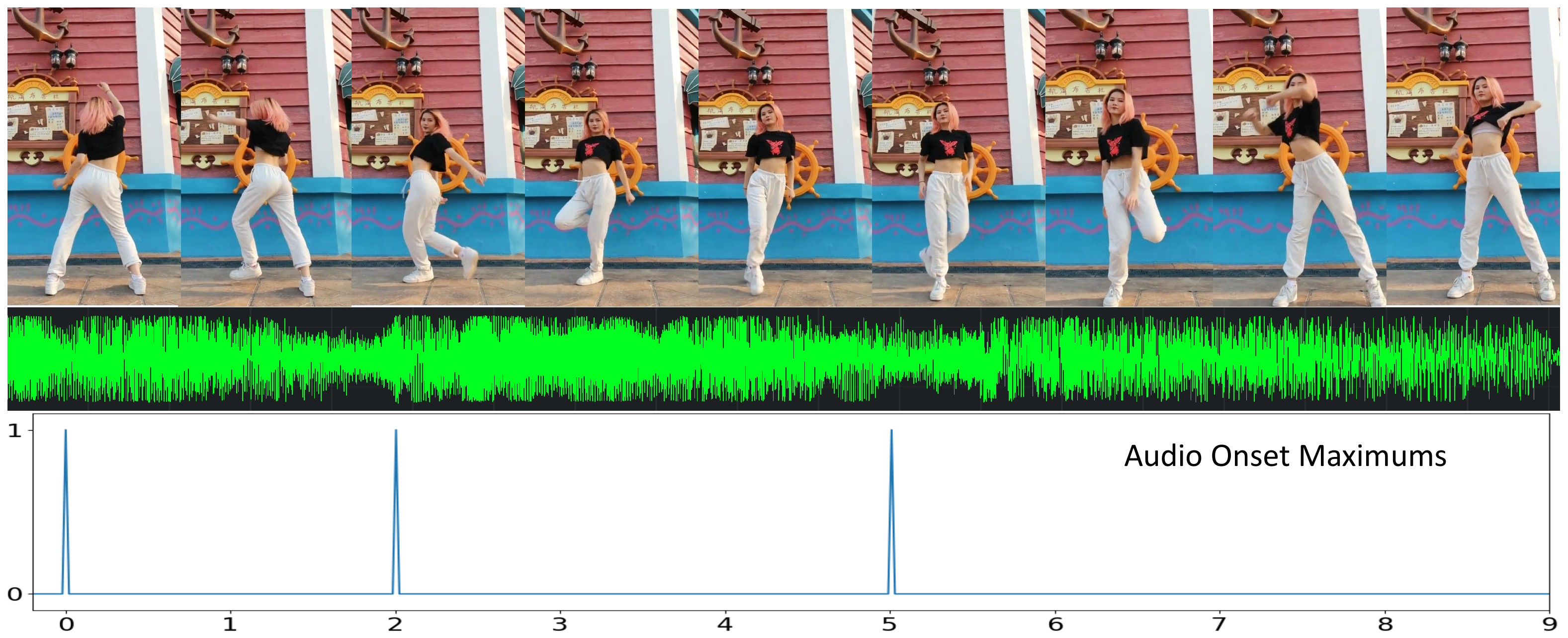}
\caption{Illustration of a dancing clip with its corresponding audio rhythms. The motion of the dancer is in sync with the music rhythms, which means the starting keypoint of the dancing motion lies on the temporal position of the music onset maximum. The correlations between motion and onset features are used as explicit training signals, which together with the implicit rhythm consistency contribute to self-supervised representation learning.}
\label{figure1}
\end{figure}

A more practical way is to leverage the correlation between auditory and visual contents as the proxy to train task-agnostic models without human annotation. The derived representation can further be applied to a variety of downstream tasks with minor modifications. Though recent audio-visual pre-trained models~\cite{chung2016out, arandjelovic2017look, arandjelovic2018objects,korbar2018cooperative, cheng2020look, owens2018audio, hu2019deep, ma2021active} have delivered an impressive performance, these methods are not generalizable for dancing scenarios. The reason is that the visual contents of dancing videos consist of complex motions and characteristics. Meanwhile, dancing music, the auditory counterpart, embodies various attributes like lyrics, rhythms, tempos, etc. These fine-grained features are neglected by existing self-supervised audio-visual methods, leading to the consequence that meticulous conjunctions crossing modalities are not fully probed. As a result, a novel multimodal representation learning strategy is necessitated for dancing videos.

% Insight
In this paper, we focus on the \textit{rhythm} of dancing videos, which represents the amplitude of music intensity and dancing motions. As illustrated in Fig.~\ref{figure1}, dancers tend to rhythmically move by music at the same frequency to make dances more coordinated. Such temporal key moments of dancing motions are usually in sync with music rhythms, which can be reflected by the audio onset features. Inspired by this insight, we argue that the pattern of dancing motions, termed as the \textit{visual rhythms}, should be synchronous with music rhythms. This temporal correspondence can further be utilized as the supervised signals. Specifically, the synchronization of music and dance rhythms is conducted both explicitly and implicitly. Since rhythms are implied in the auditory and visual contents, we leverage the consistency between audio and visual streams to perform implicit rhythm alignment. Moreover, the audio onset, which refers to the beginning of a musical note, can be regarded as the music rhythm. In this work, we train the model to learn visual rhythms based on appearance and motion cues, which are expected to be aligned with music rhythms in an explicit way. Finally, the entire model is unified by joint training for multimodal representation as well as visual rhythm extraction, which can further be applied to many downstream applications. We conduct experiments on three practical tasks: dance classification, music-dance retrieval, and music-dance retargeting. The comparison with other self-supervised audio-visual methods verifies the effectiveness of our framework. Our contributions are summarized as follows:

\begin{itemize}
\item We devise a novel self-supervised representation learning strategy for dancing videos, which performs the music-dance rhythm synchronization both in explicit and implicit ways.
\item We propose a joint music-dance representation and a dance rhythm extractor favorable for music-dance understanding and re-creation tasks.
\item Extensive experiments on three downstream tasks, i.e., dance classification, music-dance retrieval, and dance-music retargeting, verify the effectiveness and generalizability of our model on music-dance scenarios.
\end{itemize}

\section{Related Works}
\label{sec:relat}

\noindent\textbf{Audio-visual representation learning} aims to learn the multisensory embeddings favorable for downstream tasks. Several existing methods~\cite{chung2016out, arandjelovic2017look, arandjelovic2018objects,korbar2018cooperative, cheng2020look, owens2018audio, alayrac2020self, Afouras20b, morgado2021audio} focus on the characteristics of auditory and visual streams in a given video. The audio and visual contents in videos are usually corresponding and synchronous, and these natural inter-modality correlations can be utilized as the supervisory signal for large-scale self-supervised training. Apart from such correspondence, some works investigate the discriminative information across the modalities. \cite{hu2019deep, alwassel2020self} explore the intrinsic differences between modality-aware semantics and propose a cross-modal deep clustering method to perform cross-modality supervision. \cite{ma2021active} propose a cross-modal active contrastive coding strategy to fully explore the diversity between positive and negative samples. Moreover, some methods~\cite{barzelay2007harmony, Zhao_2018_ECCV, zhao2019sound, hu2020discriminative, gan2019self, gan2020music, morgado2020learning, afouras2021self} explore the relationship of visual dynamic motions and audio signals. The correspondence between the dynamic motions of objects and sound sources can be used for self-supervised object detection. In this paper, we propose a unified framework to combine these knowledge priors above, where the temporal correspondence is utilized for implicit rhythm synchronization, and the dynamic motions corresponding to music are exploited for explicit rhythm keypoint alignment. 

\noindent\textbf{Dance with music} is a fundamental and challenging video category, where the movements of dancers are complex and the patterns of music and dance are diverse. Hence, models are required to explore more detailed information by capturing fine-grained features. Attempts in the representation learning field focus on the uni-modal music embeddings~\cite{wu2021exploring, liang2020pirhdy, zhu2021musicbert, zeng2021musicbert, zhang2021dance, ye2020choreonet}. Since music can be handled as a sequence of tokens, an ordinary pipeline is to train pretrained models that have achieved promising performance in natural language processing, such as BERT~\cite{devlin2018bert}. However, the embeddings of the visual counterpart: dancing, as well as the joint music-dance representation that captures cross-modal music-dance interactions, are rarely investigated. For the multimodal dance-music applications, many works~\cite{zhuang2020music2dance, lee2019dancing, guo2021danceit, huang2020dance} tackle the problem of music-dance generation, that is, generating a sequence of dancing videos based on the given music. Some methods~\cite{wang2020alignnet} work on the audio-visual alignment, which tries to align mismatched dancing and music streams by learning frame-level dense correspondence. The most relevant to our work is visual rhythm and beat prediction~\cite{xie2019visual, davis2018visual, pedersoli2020dance}, where visual rhythms and beats are extracted based on dancing motions or dancer skeletons. In our paper, we ameliorate visual rhythm predictions by involving more characteristics and further leverage the synchronization of rhythms as a training proxy for multimodal representation learning.

\begin{figure*}[t]
\centering
\includegraphics[width=0.98\textwidth]{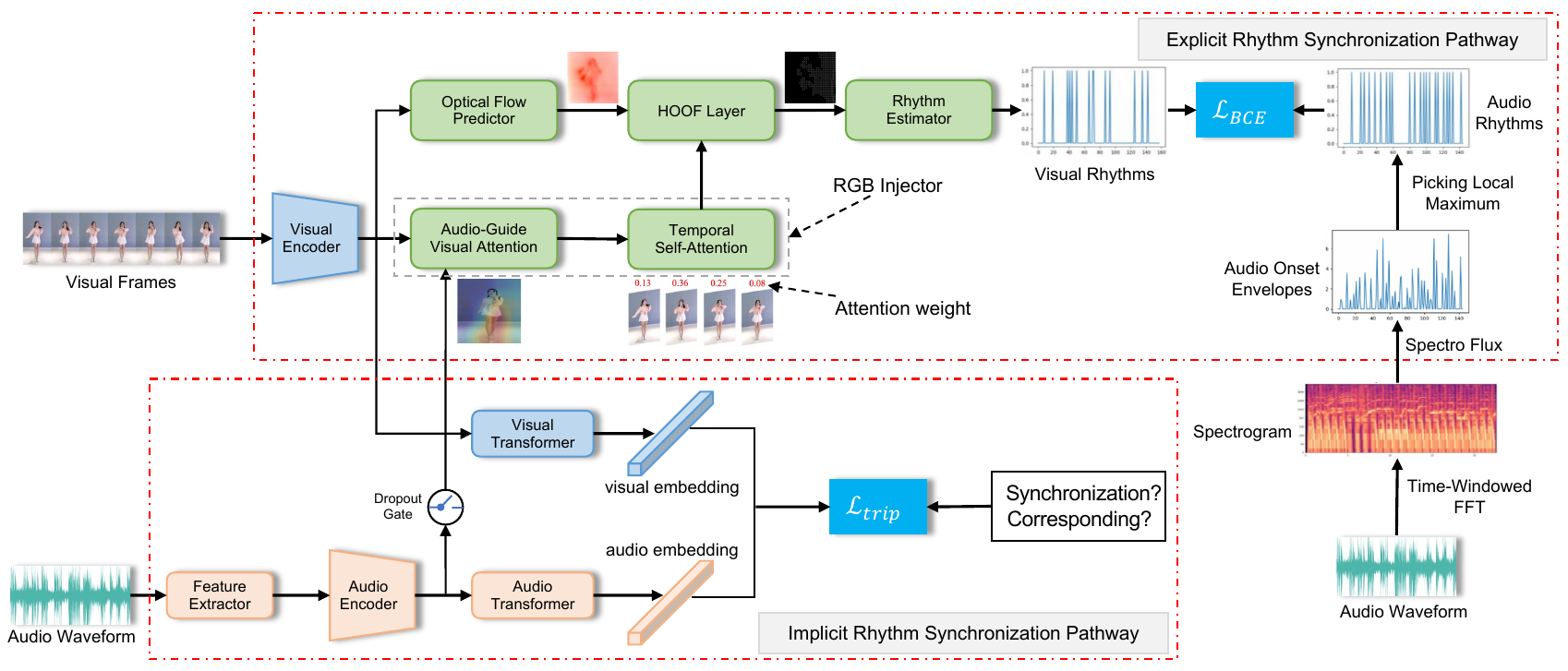}
\caption{Illustration of our MuDaR framework. MuDaR involves two pathways: implicit and explicit music-dance synchronization. For the implicit pathway, the auditory and visual features are used to conduct corresponding and synchronization predictions. For the other part, MuDaR explicitly learns visual rhythms based on the motions of dancers by temporally aligning with audio onsets, which can be viewed as music rhythms. Two streams are trained jointly in a self-supervised manner.}
\label{figure2}
\end{figure*}

\section{Approach}
\label{sec:appro}

As shown in~ Fig.\ref{figure2}, MuDaR conducts rhythm synchronization via a two-pathway architecture. The explicit pathway temporally aligns the music and dance rhythm keypoints, while the implicit part is implemented by identifying the synchrony between audio and visual streams. 

\subsection{Explicit Rhythm Synchronization}

\subsubsection{Music Rhythm Detection} 

For the music rhythms, a straightforward idea is to utilize the onset feature, which indicates sudden increases in music volumes. Following~\cite{bock2013maximum}, we first obtain the spectrograms by conducting time-windowed FFT to the audio signal:
\begin{equation}
\small
    X(n, k) = \sum_{q=-\frac{N}{2}}^{\frac{N}{2}-1} v(hn+q)w(q)e^{-\frac{2j\pi qk}{N}},
\end{equation}
\noindent where $X(n, k)$ denotes the $k^{th}$ frequency bin of time $n$; $v(\cdot)$ is the audio signal; $h$ denotes the hop size; $w(\cdot)$ is the Hamming window, $N$ is the time window size.

Then we compute the bin-wise difference between spectrograms to get the spectral flux, which indicates the magnitude of audio signals. The onset envelope, a positive 1D feature, is computed by summing the positive spectral flux:
\begin{equation}
\small
    OE(n) = \sum_{k=-\frac{N}{2}}^{\frac{N}{2}-1} max(0, \left\vert X(n, k) \right\vert - \left\vert X_{ref}(n-\mu, k) \right\vert),
\end{equation}
\noindent where $n, k$ denotes the temporal position and bin number, respectively; $OE(n)$ is the onset envelope at time step $n$; $X_{ref}$ denotes the maximum-filtered spectrogram proposed in~\cite{bock2013maximum}; $\mu$ is the time lag.

Finally, we pick the local maximum of the onset envelope to detect the discrete onset feature. The local maximum will be selected as an onset only if it is some threshold above the local average value. The entire rhythm extraction procedure can be derived by the Python package LibROSA~\cite{mcfee2015librosa} with the default hyperparameter settings.

\subsubsection{Dance Rhythm Detection} 

\noindent\textbf{Optical flow} indicates the direction and magnitudes of visual motions. Therefore, we estimate the dense optical flows as the initial step of dance rhythm detection. The lightweight PWC-Net~\cite{sun2018pwc} is selected as the backbone optical flow network. To be specific, PWC-Net extracts feature maps in different resolutions via a feature pyramid extractor consisting of several stacking convolution layers, then warps features of the current frame toward the previous frame and constructs a cost volume. Finally, the optical flow is predicted by a multi-layer CNN estimator and refined by a dilated context network.

\noindent\textbf{Histograms of optical flow.} We utilize the Histogram of Oriented Optical Flow (HOOF)~\cite{chaudhry2009histograms}, a non-Euclidean feature, to represent motions in a non-linear manifold that is more robust for dancing scenarios. Specifically, HOOF is computed by the weighted summation of the magnitude of optical flow, where the optical flow $v(t, x, y)$ in each time-stamp is separated into $n$ bins according to its angle from the horizontal axis, represented as follows:
\begin{equation}
    H(n, k) = \sum_{x, y}M_t(x, y)\mathbbm{1}_\theta (P_t(x, y)),
\end{equation}
\begin{equation}
    \mathbbm{1}_\theta(\phi) := 
    \begin{cases}
    1, &\text{if } |\theta-\phi|\le \frac{2\pi}{B}, \\
    0, &\text{otherwise},
    \end{cases}
\end{equation}
\noindent where $M_t(x, y)\in \mathbb{R}^{H\times W}$ denotes the magnitude of optical flow in $t^{th}$ time, which is computed by $\sqrt{x^2+y^2}$; $P(t)\in \mathbb{R}^{H\times W}$ indicating the angle of optical flow with x-axis in time step $t$ is calculated by $tan^{-1}\frac{y}{x}$, $\mathbbm{1}_\theta(\phi)$ is an indicator function; $B$ denotes the number of bins.

\noindent\textbf{RGB injector.} Directly using HOOF for rhythm estimation could lead the prediction to be highly dependent on the quality of optical flow. Considering that the RGB features may also contain contributing information, we propose an RGB injector to infuse visual cues to motion features as an enhancement. To be specific, we use the feature map $f_{rgb}$ generated by the encoder of PWC-Net in the explicit rhythm synchronization pathway. Then $f_{rgb}$ is tiled and linearly projected to reduce the dimensionality. To highlight the rhythm keypoints in the temporal dimension, we introduce the long-range temporal interactions by the multi-head self-attention mechanism~\cite{vaswani2017attention}. The refined visual features are put into the visual rhythm estimator as the infusion.

We also argue that the inherent coherence of music-dance signals can be utilized for visual rhythm prediction. The audio-guided spatial-channel attention mechanism~\cite{xu2020cross} is leveraged to explore the relationship between auditory and visual features. We use audio features $f_a$ extracted by the audio encoder of the implicit rhythm synchronization pathway (which will be introduced in Sec.~\ref{subsec:impli_sync}), and the audio-guided visual features can be computed as follows,
\begin{gather}
\small
    w_{a:rgb}^c = \sigma (W_1U_1^c(\rho_a(f_a \odot U_{rgb}^c(f_{rgb})))), \\
    f_{a:rgb}^c = \sum_{i=1}^k w_{a:rgb}^{c:i} f_{rgb}^i,\\
    w_{a:rgb}^s = softmax(\delta(W_2(U_a^s(f_a) \odot U_{rgb}^s(f_{rgb}^c)))),\\
    f_{a:rgb} = \sum_{i=1}^k w_{a:rgb}^{s:i} f_{rgb}^{c:i},
\end{gather}
\noindent where $U_{rgb}^c, U_{rgb}$ are linear layers with non-linearity activation; $W_1, W_2$ are learnable parameters; $\sigma$ indicate the sigmoid function; $\rho$ denotes global average pooling; $\delta$ is the hyperbolic tangent function; $w_{a:rgb}^c, w_{a:rgb}^s$ are the channel and spatial attention map, respectively.

However, the prediction of dance rhythms cannot rely on the corresponding music when applied to downstream tasks. The performance significantly declines when audio signals are available during training while missing in inference. To this end, we refer to Dropout~\cite{srivastava2014dropout}, a simple yet effective method which initial goal is to prevent the model from being dependent on specific neurons. Inspired by this paradigm, we propose an audio dropout gate, which randomly drops partial auditory inputs in a mini-batch with constant ratios during training. By doing so, MuDaR can detect visual rhythms both with and without music during inference. The entire procedure can be formulated as:
% ----------------------------------
% \begin{gather}
% f_{a:rgb1} = AGVA(f_t^{rgb}\odot g, f_t^a), 
% \end{gather}
\begin{gather}
    f_{rgb1}, f_{rgb2} = AudioDropout(f_{rgb}, p),  \\
    AudioDropout(f, p) = f[b*p:], f[:b*p], \\
    f_{a:rgb1} = AGVA(f_{rgb1}, f_a),       \\
    f'_{rgb2} = Linear(Tile(f_{rgb2})),     \\ 
    f_{a:rgb} = Concat(f_{a:rgb1}, f'_{rgb2}),  \\
    f_{inj} = Att(f_{a:rgb}, f_{a:rgb}, f_{a:rgb}),
\end{gather}
\noindent where $f_{inj}$ denotes the output of the RGB injector; AGVA denotes the audio-guided visual attention introduced in Eq.(5)$\sim$(8); Att denotes the multi-head self-attention; $b$ denotes the mini-batch size; $p$ is the audio dropout rate.

\noindent\textbf{Visual rhythm estimator.} To explore the magnitude of visual appearance, we compute the first-order difference of motion and RGB features, respectively, and combine them via linear projection. Then a linear layer is used for binary classification,
\begin{equation}
    p_t^e = \sigma(W_e(U_{mot}f'_{mot} \oplus U_{inj}f'_{inj})+b_e),
\end{equation}
\noindent where $f'_{mot}, f'_{inj}$ are the first-order difference of motion and injected features; $U_{mot}, U_{inj}$ are linear layers with ReLU~\cite{nair2010rectified}; $\oplus$ denotes concatenation across the channel dimension; $W_e, b_e$ are parameters of classifier; $\sigma$ is the sigmoid function.

\subsection{Implicit Rhythm Synchronization}
\label{subsec:impli_sync}
We claim that rhythms of music and dance are implied in the auditory and visual features, and the synchronization of music and dance streams can also be considered as the implicit version of rhythm synchronization. Following~\cite{owens2018audio, cheng2020look, afouras2021self, korbar2018cooperative}, we employ Audio-Visual Correspondence (AVC) and Audio-Visual Temporal Synchronization (AVTS) as the pretext tasks for representation learning. Specifically, we leverage the raw videos as the positive samples, while creating asynchronous samples by performing temporal shifts and creating uncorrelated samples by combining visual and audio streams from different videos. Then the model is required to predict whether the assembled sample is synchronous (for AVTS) and corresponding (for AVC) or not, thereby performing self-supervised training.

One problem of this unsupervised paradigm is the construction of \textit{false-negative samples}. For the AVC task, if we randomly sample a sequence of music that is the same as the raw dancing video, the newly-assembled sample is actually music-dance corresponding. To address this problem, we compute the \textit{rhythm similarity score} of the raw and newly selected audio streams by calculating the coincidence of rhythm positions, which is formulated as follows:
\begin{equation}
\small
    s_{rhy} = \sum_{t=1}^T \left\vert O_{pos}^t-O_{neg}^t \right\vert - \alpha T, 
\end{equation}
\noindent where $s_{rhy}$ is the rhythm similarity score; $O_{pos}^t$ denotes the positive music onset features; $O_{neg}^t$ indicates the negative onsets; T denotes the length of dancing videos; $\alpha \in [0, 1]$ is a threshold hyperparameter. The negative sample will be selected only when the score is positive. For the AVTS task, the rhythms of dancing videos are sometimes periodic. Therefore, if we shift the current rhythm point exactly to another point afterward, the audio and visual rhythms will be re-aligned with certain time-lagged. To this end, we conduct additional constraints to prevent the shifted size from being the multiple of rhythm interval. Though the rhythm interval tends to be diverse, most onset peaks are concurrent with music beats, which are distributed with a constant temporal gap. In our work, the eighth note (quaver) is selected as the basic beat unit, and the shifted frames $f_{sft}$ cannot be the multiple of the frame number identical to an eighth note,
\begin{equation}
% \small
    f_{sft} \bmod (k_{fps}*\frac{60}{k_{bpm}}*\frac{1}{2}) \ne 0,
\end{equation}

\noindent where $k_{fps}$ denotes the frame sample rate; $k_{bpm}$ denotes the number of quarter notes (crotchets) per minute. Since we use the eighth note as the base beat unit, the number of beat units per minute will be $2*k_{bpm}/60$.

After performing sampling constraints, the implicit rhythm synchronization can be trained in a self-supervised manner. To perform coherent optimization, we leverage the outputs of the visual encoder in the explicit pathway as visual inputs. We also build a similar audio encoder including several 2D convolution layers. Then auditory and visual features are put into $K$ stacking transformer~\cite{vaswani2017attention} layers, respectively. We do not involve any modality interaction to make our framework generalizable for single-modality downstream tasks. Each transformer layer is constructed by the encoder part of the raw transformer architecture~\cite{vaswani2017attention}, in which features temporally interact to fully explore the long-range temporal correlation. Finally, the triplet loss~\cite{schroff2015facenet} is used for the refined auditory and visual features, which enlarges the distance between visual features and negative auditory features, while reducing the gap between visual features and positive audio features. 

\subsection{Optimization}

For implicit synchronization, training a binary classifier and adopting the binary cross-entropy loss as the learning objective is a natural choice. However, this method suffers from the difficulty of training convergence. Some prior methods~\cite{chung2016out, korbar2018cooperative} opt for the contrastive loss as a substitution. In this work, the dense optical flow estimation requires high computation costs, thus the performance will be limited when the batch size is small. In this paper, we choose the triplet loss~\cite{schroff2015facenet} with an offline negative sample strategy. Specifically, we put a pair of positive audio and visual samples into the two-stream implicit pathway. Then we choose the negative audio samples from the whole dataset, which are brought into the audio streams together with the positive audio sample. The triplet loss function tries to make the refined positive audio feature $f'_{a:pos}$ closer to the refined visual features $f'_{v}$, while enlarging the distance between negative audio features $f'_{a:neg}$ and $f'_{v}$,
\begin{equation}
\small
    \mathcal{L}^{im} = \sum_{i=1}^N \bigg[|| f'_v-f'_{a:pos}||_2^2-||f'_v - f'_{a:neg}||_2^2 + \alpha\bigg]_+,
\end{equation}

\noindent where $\mathcal{L}^{im}$ denotes the implicit loss, $\alpha$ is the margin hyperparameter; $[\cdot]_+$ denotes picking the positive results, and $N$ denotes the number of triplet pairs.

For explicit synchronization, the straightforward way of using binary cross-entropy loss may cause overwhelming negative predictions during training due to the unbalanced distribution of rhythm and non-rhythm temporal keypoints. To address this problem, we adopt the focal loss~\cite{lin2017focal}, which addresses the class imbalance by introducing a weighting hyperparameter $\alpha$, and focuses more on hard samples rather than easier ones via the focusing parameter $\gamma$. The explicit loss $\mathcal{L}^{ex}$ can be formulated as,
\begin{equation}
% \small
    \mathcal{L}^{ex} = -\alpha_t(1-p_t^e)^\gamma log(p_t^e),
\end{equation}

\noindent where $p_t$ is the visual rhythm prediction.

Finally, the optimization is conducted in a joint-training manner, the overall objective function is formulated as,
\begin{equation}
    \mathcal{L} = \lambda_1 \mathcal{L}^{im} + \lambda_2\frac{1}{N}\sum_{t=1}^N \mathcal{L}_t^{ex},
\end{equation}
\noindent where $\lambda_1, \lambda_2$ are weighted hyperparameters to control the summation of loss terms, $N$ is the frame number.

\section{Experiments}
\label{sec:exper}

\subsection{Music-Dance Representation Learning}

\noindent\textbf{Dataset.} MuDaR can be trained by large-scale dancing videos without any annotation. The only requirement for unlabeled data is that dancers should twist their bodies following the rhythms of the music, which is a fundamental rule for dancing. In this paper, we collect 194,407 dancing videos from an information feed platform and online video-sharing application. The training data can be separated into four categories: dancing with pop music, hip-hop, modern dance, and K-pop with nearly equal proportions. These category labels are unnecessary both in training and inference.

\noindent\textbf{Implementation details.} All dancing videos are cropped to 8s clips. To remove background noises, we discard the beginning and ending parts of each dance, and select 4 seconds before the middle of the video together with 4 seconds after. Visual frames are sampled with a frame rate of 8 fps. Then frames are resized to 256$\times$256. To maintain the camera view unchanged, we do not apply any visual transformation for data augmentation. Audios are sampled with a rate of 16kHz, and we extract Mel-spectrograms, onset envelopes, onset maximum, and beat using the python package LibROSA~\cite{mcfee2015librosa}. For the implicit synchronization pathway, the audio inputs are the concatenation of Mel-spectrograms, beat, and onset envelopes, while the onset maximums are used as the ground truth of rhythms for explicit synchronization. We use Adam~\cite{kingma2014adam} optimizer with the initial learning rate of 1e-4. The entire model is trained on 64 NVIDIA Tesla V100 GPUs for 50 epochs with a batch size of 384. For the implicit synchronization, we adopt the curriculum learning strategy as~\cite{korbar2018cooperative}. AVC is used for the first 35 epochs, while the more challenging pretext task AVTS is used for the rest of the training process.

\begin{table}[t]
\centering
\caption{The evaluation of visual rhythm prediction compared with two ablated models. We also report the performance of our full model with different dataset sizes.}
\label{table1}
\begin{tabular}{@{}lccc@{}}
\toprule
Approach            & Data Size & Recall & Precision \\ \midrule
MuDaR w/ BCE        & 190k                                                   & 82.2\% & 37.0\%    \\
MuDaR w/ regression & 190k                                                   & 43.8\% & 51.9\%    \\
MuDaR w/ CRF        & 190k                                                   & 70.2\% & 79.8\%    \\ \midrule
MuDaR (full)        & 47k                                                    & 59.6\% & 73.1\%    \\
MuDaR (full)        & 140k                                                   & 71.4\% & 78.9\%    \\ \midrule
MuDaR (full)        & 190k                                          & \textbf{73.2}\% & \textbf{82.0}\%    \\ \bottomrule
\end{tabular}
\end{table}

\noindent\textbf{Evaluation of MuDaR.} We randomly sample 4,407 videos of the original 194,407 videos for evaluation, while using the remaining 190k videos for self-supervised training. To fully investigate the performance of the explicit rhythm pathway, we propose three ablated models for comparison. ``MuDaR w/ CRF" denotes an ablated model that replaces the binary prediction layer with a BiLSTM-CRF~\cite{lample2016neural} layer, which is composed of a bidirectional long-short memory network~\cite{huang2015bidirectional} and a conditional random field~\cite{lafferty2001conditional} module. This ablated model regards the visual rhythm prediction task as a sequence labeling problem and considers the inner relationship of the consecutive frames. ``MuDaR w/ regression" takes the visual rhythm prediction task as a regression problem. Instead of using local maximum over onset envelopes as the audio ground truth, it directly minimizes the distance between model output and the onset envelope curve. The visual rhythms are then generated by picking the local maximum of the model output. ``MuDaR w/ BCE" replaces the binary focal loss with the raw binary cross-entropy loss. As shown in Fig.~\ref{table1}, we report recall and precision as the evaluation metrics. Results show that ``MuDaR w/ BCE" achieves higher recall compared with raw MuDaR, while obtaining low precision, proving that the cross-entropy loss suffers from the imbalanced class distribution. ``MuDaR w/ regression" achieves poor performance on all metrics. We argue that though the temporal position of music and dance rhythms are coincident, it's illogical to force the intensity of visual rhythms to be identical to the audio rhythms. On the contrary, the full MuDaR model achieves promising results with respect to all metrics, showing the effectiveness of detecting visual rhythms. ``MuDaR w/ CRF" performs slightly worse than the full model. This suggests that the BiLSTM-CRF performs unsatisfied when only two unbalanced categories exist for sequence labeling.

\subsection{Dance Classification}

\noindent\textbf{Downstream architecture.} We first evaluate the performance of MuDaR on the task of dance classification. To fully utilize the outputs of MuDaR (audio and visual embeddings, visual rhythms), we concatenate the visual embeddings and rhythms as visual outputs, and the auditory embeddings and onset as audio outputs. Then the audio and visual output features are put into a two-stream classifier. To be specific, audio and visual features are first put into two fully-connected layers with non-linear activation. Then the audio and visual features are concatenated and integrated by temporal pooling. Another two fully-connected layers with non-linearity are used for prediction.

\noindent\textbf{Dataset.} We evaluate the performance of dance classification on the Let's Dance~\cite{castro2018let} dataset. This dataset contains more than 1,400 10-second dancing videos on a variety of 16 dance categories. Since a small part of the videos is unavailable on the Internet, we downloaded 1,262 dancing videos and randomly split the training, validation, and testing set with the proportion of 80\%/10\%/10\% following~\cite{castro2018let, wysoczanska2020multimodal}. We employ the same data preprocessing procedure as the self-supervised representation training.

\noindent\textbf{Implementation details.} Adam~\cite{kingma2014adam} optimizer is used for training with the initial learning rate of 4e-3, which is degraded by 10 after 20 and 50 epochs. The model is trained with a batch size of 64 for 70 epochs.

\noindent\textbf{Experimental results.} We compare MuDaR with two supervised methods: Temporal Three-Stream CNN~\cite{castro2018let} and Multimodal Dance Recognition~\cite{wysoczanska2020multimodal}. To find out whether the dance-music representation is really needed, or whether it can be replaced by larger-scale generalized audio-visual pre-trained models, we also conduct experiments with the traditional audio-visual pre-trained models XDC~\cite{alwassel2020self} and AVID-CMA~\cite{morgado2021audio} by fine-tuning them on the dance classification dataset. Furthermore, we also re-implement three audio-visual self-supervised methods: Multisensory~\cite{owens2018audio}, AVTS~\cite{korbar2018cooperative}, and LLA~\cite{cheng2020look}, which are trained on the same large-scale dancing dataset used for MuDaR to make a fair comparison. Results shown in Tab.~\ref{table2} indicate that all audio-visual pre-trained models perform worse than our proposed framework pre-trained on the music-dance dataset, even in a far larger-scale dataset (240K, 2M, and 65M vs. 190K). We argue that models pre-trained on the generalized audio-visual dataset lack the discrimination of music and dance patterns, hence are unsuitable for the dance-music tasks. Moreover, our model outperforms all self-supervised baselines trained on the music-dance dataset by a large margin, showing the effectiveness of our proposed framework in tackling dance-music tasks. Last but not least, the performance on the full pre-trained dataset (190k) surpasses the state-of-the-art supervised method, which indicates the rationality of our self-supervised paradigm. We also conduct ablation studies by removing the visual and auditory rhythms. Results show that MuDaR performs slightly worse without rhythm involvement. However, the decline is not significant, revealing that the performance of dance classification mainly depends on the embeddings generated by the implicit synchronization pathway. 

\begin{table}[t]
\centering
\caption{The dance classification accuracy on the Let's Dance dataset. ``S-Sup." denotes models pretrained on the dataset specified in the second column, then fine-tuned on the dance classification dataset. M\&D denotes the music-dance dataset.}
\label{table2}
\begin{tabular}{@{}lccc@{}}
\toprule
Approach                                & Dataset                        & Acc. & Manner                \\ \midrule
Castro et al.~\cite{castro2018let}      & /                              & 70.2\%   & Supervised                  \\
MDR~\cite{wysoczanska2020multimodal}    & /                              & 77.0\%   & Supervised                  \\ \midrule
LLA~\cite{cheng2020look}                & AudioSet (240k)                 & 66.9\%   & S-Sup.+ ft                 \\ 
XDC~\cite{alwassel2020self}             & AudioSet (2M)                   & 71.1\%   & S-Sup.+ ft                \\
XDC~\cite{alwassel2020self}             & IG-Kinetics (65M)               & 73.2\%   & S-Sup.+ ft            \\ 
AVID-CMA~\cite{morgado2021audio}        & AudioSet (2M)                   & 75.9\%   & S-Sup.+ ft            \\  \midrule
Multisensory~\cite{owens2018audio}      & M\&D (190k)                      & 71.4\%   & S-Sup.+ ft            \\
AVTS~\cite{korbar2018cooperative}       & M\&D (190k)                      & 68.3\%   & S-Sup.+ ft            \\
LLA~\cite{cheng2020look}                & M\&D (190k)                      & 73.0\%   & S-Sup.+ ft            \\ \midrule
MuDaR (full)                            & M\&D (47k)                       & 76.1\%   & S-Sup.+ ft            \\
MuDaR (full)                            & M\&D (140k)                      & 79.4\%   & S-Sup.+ ft            \\
MuDaR w/o rhy                           & M\&D (190k)                      & 80.8\%   & S-Sup.+ ft            \\ \midrule
MuDaR (full)                            & M\&D (190k)                      & \textbf{81.7}\%   & S-Sup.+ ft   \\ \bottomrule
\end{tabular}
\end{table}

\subsection{Music-Dance Retrieval}

\noindent\textbf{Downstream architecture.} We also evaluate our model on the music-dance retrieval task. For the cross-modal retrieval, we compute the similarity scores of rhythms and embeddings, respectively. To be specific, for the embedding $E_a, E_v$ and rhythm $R_a, R_v$, we compute the embedding similarity matrix $S_e$ consisting of similarity scores of $E_a$ and $E_v$, then compute the rhythm similarity matrix $S_r$ via $R_a$ and $R_v$. Finally, the hybrid similarity matrix can be computed by the weighted summation of $S_e$ and $S_r$:
\begin{equation}
    S_{hyb} = \lambda_3 S_e + (1-\lambda_3) S_r,
\end{equation}

\noindent where $\lambda_3$ is the hyperparameter. The top-K indices of the hybrid matrix are preserved as the retrieval result.

\noindent\textbf{Dataset and task formulation.} We conduct experiments on the Dance-50~\cite{wang2020alignnet} dataset for cross-modal retrieval. Dance-50 contains 50 hours of K-pop dancing videos collected from online video platforms. This dataset is originally used for audio-visual alignment without any category annotation available. To perform music-dance retrieval, we selected 22 pieces of dancing music that appeared with high frequency, where each song corresponds to 15-40 dancing videos from different dancers. We collected 400 labeled videos in total while using another 1833 unlabeled dances as irrelevant data in the retrieval database. Since dances in the Dance-50 dataset are long videos of more than 30 seconds, we pre-process each annotated video by clipping 24 seconds from the beginning. For the unlabelled videos, we separated each video into several 24s dance clips. In this way, we got 400 labeled and 4,066 unlabeled dancing clips as the retrieval dataset. The objective of this task is to retrieve related dances given the targeted music.

\noindent\textbf{Experimental results.} We leverage the self-supervised methods Multisensory~\cite{owens2018audio}, AVTS~\cite{korbar2018cooperative}, and LLA~\cite{cheng2020look} pre-trained on music-dance dataset, and the available XDC~\cite{alwassel2020self} and AVID-CMA~\cite{morgado2021audio} models pre-trained on traditional audio-visual datasets as baselines. For these baselines, since visual rhythms are unavailable, we only use the similarity matrix between embeddings for retrieval. We also investigate the role of visual rhythms via the ablated model ``MuDaR w/o rhy", and we report average top-k retrieval performance (R@K) and top-k precision (P@K) as the evaluation metrics. As shown in~ Tab.\ref{table3}, results first reveal that on the music-dance retrieval task, where the downstream task is fulfilled in an unsupervised way, all traditional audio-visual pre-trained models show poor performance and are incapable of retrieving satisfying results due to the semantically similar nature of different dances and songs. Subsequently, MuDaR outperforms all music-dance pre-trained models by a large margin. Especially, our framework outperforms baseline LLA by 15\% on R@5 and P@10, and 13\% on R@1, which shows the effectiveness on unsupervised downstream tasks. Besides, the performance significantly declines after removing the rhythm information, indicating that rhythms are favorable for cross-modal retrieval.

\begin{table}[t]
\centering
\caption{Evaluation of music-dance retrieval compared with other self-supervised methods. MuDaR w/o rhy denotes the ablated model without rhythm involvement. M\&D denotes the music-dance dataset.}
\label{table3}
\begin{tabular}{@{}lcccc@{}}
\toprule
Approach                               & Dataset                  & R@1      & R@5      & P@10      \\ \midrule
LLA~\cite{cheng2020look}               & AudioSet (240k)           & 0.105    & 0.287    & 0.112     \\ 
XDC~\cite{alwassel2020self}            & AudioSet (2M)             & 0.147    & 0.219    & 0.180     \\
XDC~\cite{alwassel2020self}            & IG-Kinetics (65M)         & 0.194    & 0.315    & 0.221     \\ 
AVID-CMA~\cite{morgado2021audio}       & AudioSet (2M)             & 0.162    & 0.267    & 0.208     \\  \midrule
Multisensory~\cite{owens2018audio}     & M\&D (190k)               & 0.468    & 0.732    & 0.480     \\
AVTS~\cite{korbar2018cooperative}      & M\&D (190k)               & 0.430    & 0.698    & 0.467      \\
LLA~\cite{cheng2020look}               & M\&D (190k)               & 0.501    & 0.781    & 0.512      \\ \midrule
MuDaR (full)                           & M\&D (47k)                & 0.532    & 0.782    & 0.596      \\
MuDaR (full)                           & M\&D (140k)               & 0.604    & 0.894    & 0.632       \\
MuDaR w/o rhy                          & M\&D (190k)               & 0.586    & 0.872    & 0.615         \\ \midrule
MuDaR (full)                           & M\&D (190k)               & \textbf{0.622}    & \textbf{0.924}    & \textbf{0.661}        \\ \bottomrule
\end{tabular}
\end{table}

\subsection{Music-Dance Retargeting}

\noindent\textbf{Task formulation.} Music-dance retargeting aims to synthesize a new dancing video via two mismatched music and dance clips. This task can be combined with the music-dance retrieval task as an automatic soundtrack generator, which synthesizes rhythm-aligned dancing videos after obtaining related music via cross-modal retrieval.  Concretely, given a sequence of visual frames $V$, and a music clip $A$ from different dancing videos, the objective of music-dance retargeting is to combine $V$ and $A$ into a natural new dancing video. This task evaluates the performance of rhythm extraction thoroughly since we conduct the retargeting by warping the visual rhythms $R_v$ into the alignment with the auditory counterpart $R_a$. 

\noindent\textbf{Downstream architecture.} Due to the difference between the temporal lengths of $V$ and $A$, we try to align as many rhythm points as possible via acceleration and shifting. We propose three retargeting patterns: temporal shifting, temporal acceleration, and dynamic time warping. For the temporal shifting, we propose a sliding window, which size is equal to the length of the shorter sequence, and put it on the rhythm of the longer duration. We reserve the cropped clip with the most similar number of rhythm points, then align the starting rhythm points of two sequences. The second pattern is temporal acceleration, where we compute the temporal interval $I$ between two adjacent rhythm points: $I_k^m = P_{k+1}^m - P_{k}^m$, where $m\in \{a, v\}$ denotes the modality index, $k$ indicates the rhythm keypoint index, and $P_k^m$ denotes the temporal position of the $k^{th}$ rhythm in the modality $m$. Then we manage to make $I_k^v$ and $I_k^a$ equal for all rhythm indices by interpolating and removing visual frames. The third manner is dynamic time warping (DTW)~\cite{berndt1994using, muller2007dynamic}, which is capable of selecting the most suitable aligning strategy. Specifically, DTW aims to find the optimal warping path that aligns $R_v$ and $R_a$ with minimum rhythm mismatch cost, which is defined by the accumulated distance between the rhythm keypoints. The cumulative distance $c$ in position $(i, j)$ can be computed using dynamic programming:
\begin{equation}
\small
    c(i, j) = d(i, j) + M\{c(i-1, j-1), c(i-1, j), c(i, j-1)\},
\end{equation}
\begin{equation}
\small
    d(i, j) = |P_i-P_j|,
\end{equation}

\noindent where $M(\cdot)$ denotes the minimize function; $i, j$ are indices of rhythms; $P_i$ denotes the temporal position of the $i^{th}$ rhythm point. The warping path determines the detailed frame accelerate strategy with minimum warping costs, thereby resulting in optimal retargeting performance.

\noindent\textbf{User study.} Since there is no feasible quantitative metric for music-dance retargeting, we conduct a user study to compare our method with the only available baseline method visbeat~\cite{davis2018visual}. We survey 20 experienced dancing viewers, which are required to give judgments to 10 retargeted demos about synthesized fluency, music-dance rhythm consistency, and viewing naturalness. Results shown in Table~\ref{table:quanti} show that our model outperforms the baseline method by a large margin, indicating that our proposed method is capable of generating high-quality retargeted video-music pairs.

\begin{table}[htbp]
\centering
\caption{User study on the music-dance retargeting task. Raw denotes directly combining the given audio with the raw video.}
\label{table:quanti}
\begin{tabular}{@{}lcccc@{}}
\toprule
Method          & Fluency   & Naturalness   & Consistency        & Average                      \\ \midrule
Raw             & 7.75      & 5.25          & 5.40               & 6.13                         \\
visbeat~\cite{davis2018visual}              & 5.65      & 5.22          & 6.70               & 5.86                         \\
Ours             & 8.10      & 7.10          & 8.00               & \textbf{7.73}                \\ \bottomrule
\end{tabular}
\end{table}

\begin{figure*}[htbp]
\centering
\includegraphics[width=0.97\textwidth]{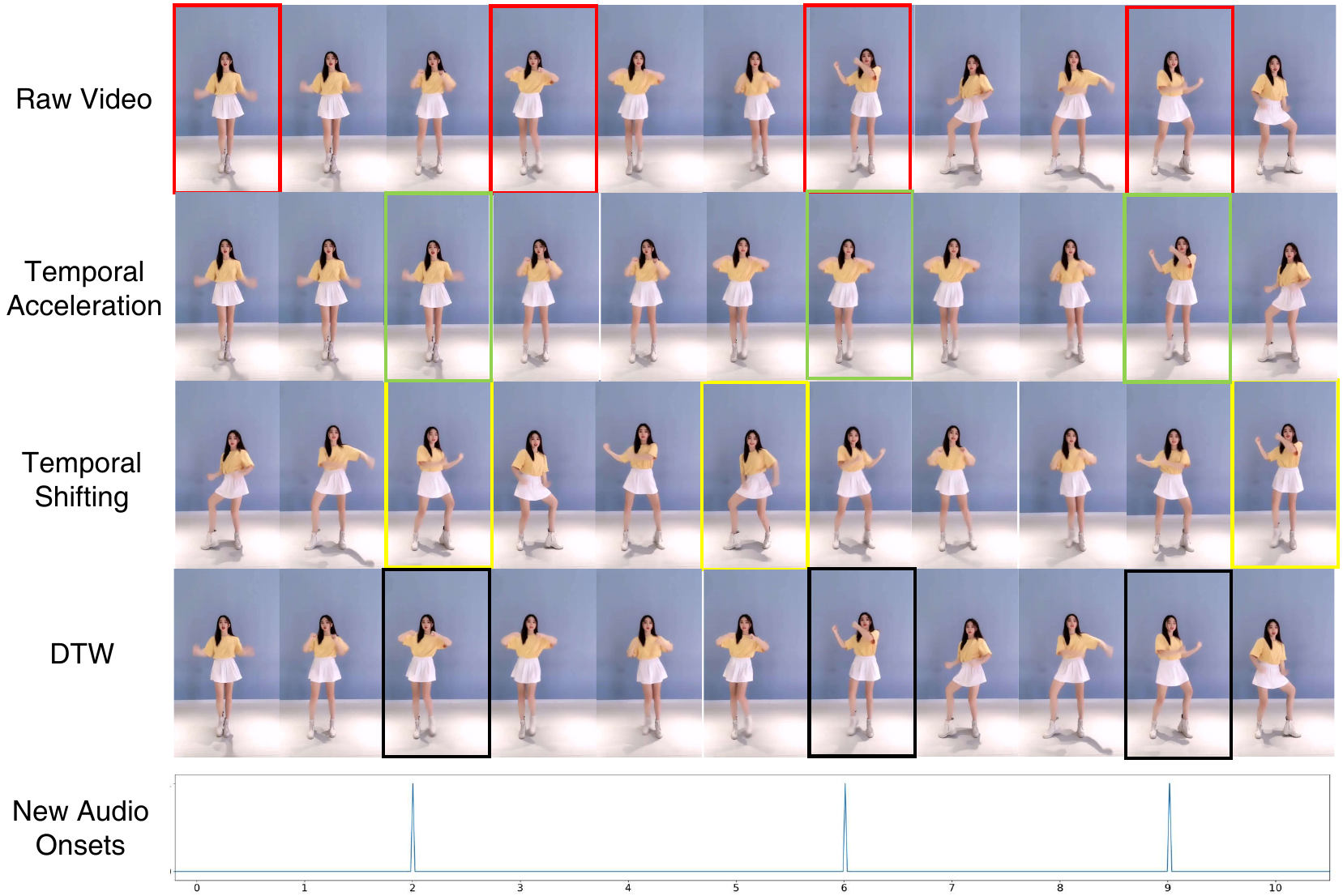}
\caption{Qualitative results on the music-dance retargeting task. Frames with colored boxes denote visual rhythm positions.} We compare the raw videos with retargeted videos via temporal acceleration, temporal shifting, and dynamic time warping (DTW). Results show that videos synthesized by DTW achieve both optimal rhythm alignment and viewing fluency.
\label{figure3}
\end{figure*}

\noindent\textbf{Qualitative results.} We also provide some qualitative results to show the performance of our method as shown in Fig.~\ref{figure3}. We compare the raw videos with retargeted videos via temporal acceleration, temporal shifting, and dynamic time warping (DTW), where frames with colored boxes denote visual rhythm positions. All dancing videos are randomly selected from the test set of the representation learning dataset. For the synthetic videos by temporal shifting, the inner rhythm may be mismatched for we only perform alignment on the first rhythm points, which makes the synthetic video uncoordinated. Retargeting videos by temporal acceleration lead parts of the synthetic video extremely fast or slow. This is because when the gap between $I_k^v$ and $I_k^a$ is quite large, we need to interpolate or remove a large number of visual frames, thereby seriously affecting viewing fluency. On the contrary, the synthetic video by DTW achieves excellent rhythm alignment compared with videos generated by temporal shifting, and the synthetic frames are more consistent than the temporal acceleration production. To make a qualitative comparison, we also provide some retargeted demos in the supplementary material synthesized by MuDaR and visbeat~\cite{davis2018visual}, which detects visual rhythms in a non-deep-learning based manner and retargets via temporal warping. Results show that the synthesized video using MuDaR and our DTW strategy achieves better viewing fluency and rhythm correspondence than the compared baselines. We strongly suggest readers refer to the provided demos since the qualitative comparison is the only possible way to evaluate the performance of our method.

\subsection{Ablation Studies}

\noindent\textbf{Ablations on MuDaR architecture.} We conduct more ablation experiments to investigate the effectiveness of different components, and the results are shown in Tab.~\ref{appen:table1}. We propose several ablated models for music-dance representation learning. ``MuDaR w/o injector" denotes removing the entire RGB injector module, `MuDaR w/o audio-guide" denotes RGB injector that only consists of the temporal self-attention. ``MuDaR w/o sa" is an ablated model where the temporal self-attention mechanism is removed. `MuDaR w/o HOOF" indicates replacing the HOOF layer with linear projection blocks, which transforms the original optical flow to the features with identical dimensions as the output of the HOOF layer. Results show that our full model outperforms ``MuDaR w/o injector" and ``MuDaR w/o HOOF" by a large margin, suggesting the necessity of the HOOF layer and the RGB 
feature infusion. The considerable performance gap between the full model and ``MuDaR w/o sa" and ``MuDaR w/o audio-guide" also proves the advantage of our proposed modules.

\begin{table}[htbp]
\centering
\caption{Ablation studies on the MuDaR architectures.}
\label{appen:table1}
\begin{tabular}{@{}lcc@{}}
\toprule
Approach                    & Recall & Precision \\ \midrule
MuDaR w/o injector          & 64.3\% & 70.8\%    \\
MuDaR w/o sa                & 71.4\% & 78.0\%    \\
MuDaR w/o audio-guide       & 70.2\% & 75.5\%    \\
MuDaR w/o HOOF              & 56.8\% & 67.2\%    \\ \midrule
MuDaR (\textbf{full})                & \textbf{73.2}\% & \textbf{82.0}\%    \\ \bottomrule
\end{tabular}
\end{table}

\begin{table}[htbp]
\centering
\caption{Ablation studies on the impact of audio dropout gate $p$.}
\label{appen:table2}
\begin{tabular}{@{}ccc@{}}
\toprule
Dropout Rate                & Recall & Precision \\ \midrule
0                           & 71.2\% & 77.5\%    \\
0.3                         & 71.8\% & 78.9\%    \\
0.7                         & 70.9\% & 76.0\%    \\
1                           & 68.1\% & 74.7\%    \\ \midrule
0.5 (\textbf{ours})         & \textbf{73.2}\% & \textbf{82.0}\%    \\ \bottomrule
\end{tabular}
\end{table}

\noindent\textbf{Impact of audio dropout gate.} We also conduct additional ablation studies on the audio dropout rate $p$. We set the audio dropout rate to different values during training, while only using visual frames as model input for inference. As shown in Tab.~\ref{appen:table2}, our full model performs slightly better compared with models with $p=0.3$ and $p=0.7$. Moreover, the performance significantly declines when setting $p=1$ due to the unbalanced modality distribution between the training and inference procedure.

\begin{table}[htbp]
\centering
\caption{Ablation studies on the impact of different transformer architecture.}
\label{appen:table3}
\begin{tabular}{@{}lcc@{}}
\toprule
Approach                             & Classification Accuracy            \\ \midrule
MuDaR w/ conv                        & 76.5\%              \\ 
MuDaR w/ trans*3                     & 79.8\%              \\ 
MuDaR w/ trans*12                    & 80.6\%              \\ \midrule
MuDaR (\textbf{full})                & \textbf{81.7}\%     \\ \bottomrule
\end{tabular}
\end{table}

\noindent\textbf{Impact of Transformer architecture.} To fully investigate the performance of the transformer architecture, we perform ablation studies on the dance classification tasks with different implicit synchronization settings. ``MuDaR w/ conv" denotes replacing the temporal convolution layers to the convolution layers used in~\cite{korbar2018cooperative}. ``MuDaR w/ trans*3" and ``MuDaR w/ trans*12" denotes stacking a different number of transformer layers. As shown in Tab.~\ref{appen:table3}, our model outperforms ``MuDaR w/ conv" by 5.2\%, showing that the transformer-based architecture achieves better results than the convolutional network. {The comparisons among transformer-based architectures show that ``MuDaR w/ trans*12" perform worse than our 6-layer version. We argue that the parameter amount of our 6-layer model is enough to serve the dance classification problem, and a larger pre-trained model may lead to the under-fit over the downstream dataset. This also suggests that straightforwardly introducing parameters cannot bring more benefits, and setting layer number 6 balances the classification performance and compute complexity.

\noindent\textbf{Ablations on training objectives.} As we mentioned above, using InfoNCE loss with a large batch can introduce more negative samples, which brings better performance yet also lead to a large training cost. When training on a Tesla V100 GPU, only 3 video-music pairs can be put into a mini-batch, which means the selection of negative samples is highly limited if we adopt an online negative-picking strategy. We also conduct ablation studies to use the contrastive InfoNCE as training objectives, and the results are shown in Table~\ref{table:loss}, which indicates that using InfoNCE with a small batch size results in poor performance.

\begin{table}[t]
\centering
\caption{Ablation studies on the impact of different training objectives.}
\label{table:loss}
\begin{tabular}{@{}lccc@{}}
\toprule
Approach            & Data Size & Recall & Precision \\ \midrule
MuDaR w/ InfoNCE        & 190k                                                   & 54.3\% & 67.1\%    \\ \midrule
MuDaR (full)        & 190k                                          & \textbf{73.2}\% & \textbf{82.0}\%    \\ \bottomrule
\end{tabular}
\end{table}

\noindent\textbf{Ablations on lambda parameters.} We also provide additional experiments of the weighted hyperparameter $\lambda_1, \lambda_2$ in Tab~\ref{table2}, where results show that adopting $\lambda_1=1. \lambda_2=5$ achieves the optimal performance.

\begin{table}[htbp]
\centering
\caption{Ablation studies on the weighted hyperparameter $\lambda_1, \lambda_2$.}
\label{table2}
\begin{tabular}{@{}lccccc@{}}
\toprule
$\lambda_1+\lambda_2$          & 1+1 & 1+3 & 1+5 & 3+1 & 5+1 \\ \midrule
Recall          & 69.0\% & 71.9\% & \textbf{73.2\%} & 68.4\% & 64.9\% \\
Precision       & 78.6\% & 79.4\% & \textbf{82.0\%} & 80.9\% & 87.2\% \\ \midrule
\end{tabular}
\end{table}

\section{Limitation and Future Works}
MuDaR has two main limitations. One is that the dense flow estimation requires high computation costs and GPU usage, further resulting in slow running speed. Moreover, the high memory utilization restricts the batch size of a single GPU, which further limits the performance of self-supervised contrastive learning since the number of negative samples is crucial for the online negative sample strategy. Another limitation is that the explicit visual rhythm prediction module, the major component of MuDaR, is highly relied on the performance of optical flow estimation. This leads the model to behave unsatisfactorily in some specific scenarios, such as multi-dancer videos, where optical flows can be extremely messy. Therefore, taking more visual features, such as motion trajectory, human gesture, and human skeleton into consideration could be a promising way to enhance the robustness and generalization of our model.

In this paper, we conduct experiments on three downstream tasks. However, we argue that MuDaR is capable of generalizing to more dance-music applications, such as music-dance realignment~\cite{wang2020alignnet} and music-inspired dance synthesis~\cite{guo2021danceit}. Visual rhythm features can also be utilized to uni-modal dancing tasks, including visual rhythm prediction~\cite{xie2019visual}, automatic dance scoring, etc. We will further extend our model to more application scenarios.

\section{Conclusion}
\label{sec:concl}

In this paper, we propose a novel self-supervised \textbf{Mu}sic-\textbf{Da}nce \textbf{R}epresentation learning framework via the synchronization of music and dance rhythms both explicitly and implicitly. We first explicitly extract and align the visual and auditory rhythms of the dancing videos based on the magnitudes of dancer motions and music contents. This part of the framework can also be utilized as a dancing rhythm extractor. We then leverage the contrastive learning strategy to synchronize the auditory and visual streams of the dancing videos, which can also be viewed as an implicit music-dance rhythm synchronization. Our model outperforms other self-supervised methods in downstream tasks by a large margin, verifying the effectiveness of our framework from both the video understanding and re-creation perspectives. We hope our work could arouse the gorgeous growth of works on music-dance self-supervised learning, and we believe the paradigm and benchmarks we proposed can contribute to more significant dance-music research.

\section{acknowledgement}
This work was supported by National Natural Science Foundation of China (No. 62172101 ).

\ifCLASSOPTIONcaptionsoff
  \newpage
\fi



\bibliographystyle{IEEEtran}
\bibliography{reference.bib}


% \end{thebibliography}

\begin{IEEEbiography}[{\includegraphics[width=1in,height=1.25in,clip,keepaspectratio]{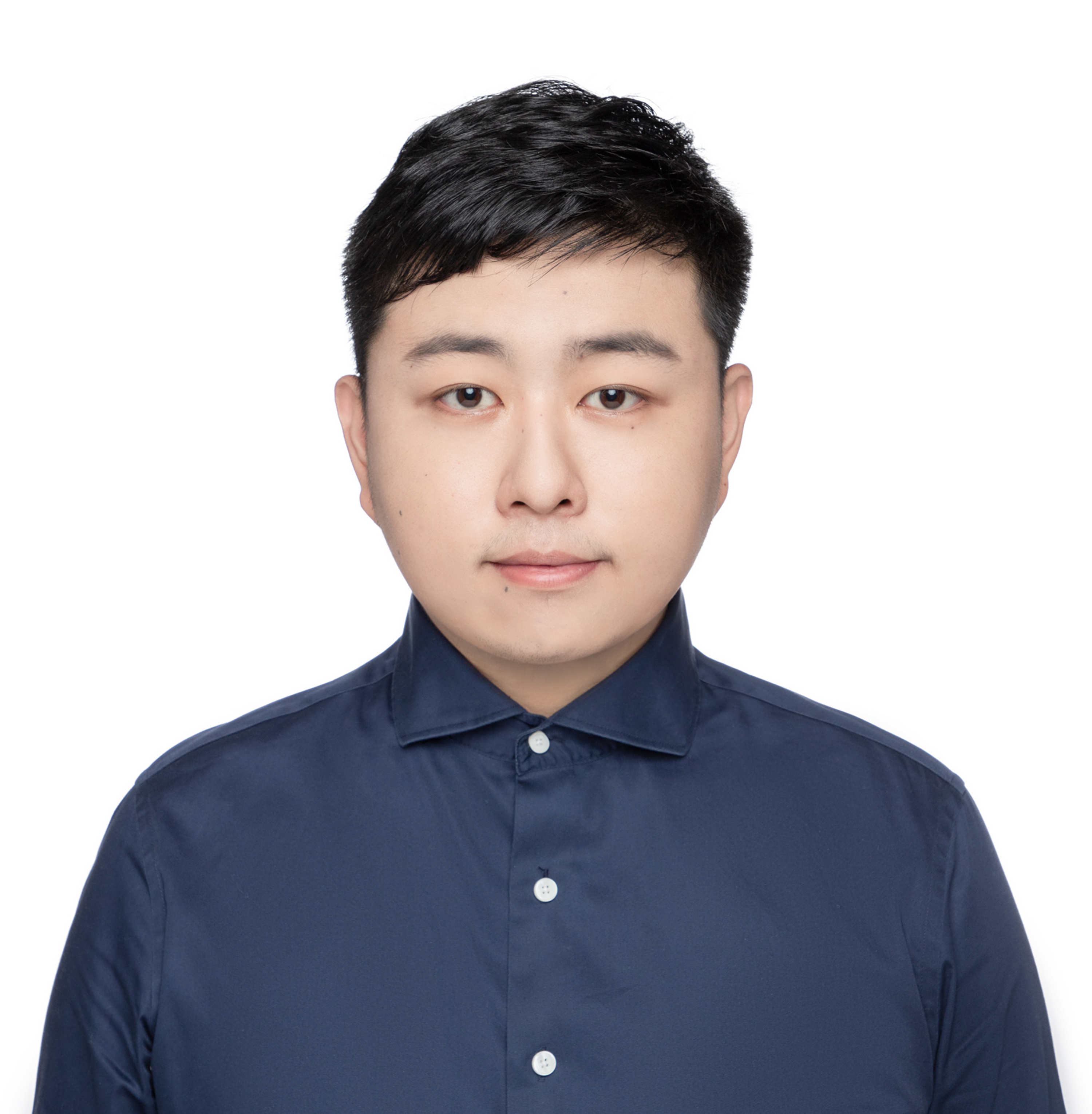}}]{Jiashuo Yu} is a master student in computer technology at School of Computer Science, Fudan University. His research interest includes audio-visual learning, self-supervised learning, multi-modality, AIGC, and Music AI.
\end{IEEEbiography}

\begin{IEEEbiography}
[{\includegraphics[width=1in,height=1.25in,clip,keepaspectratio]{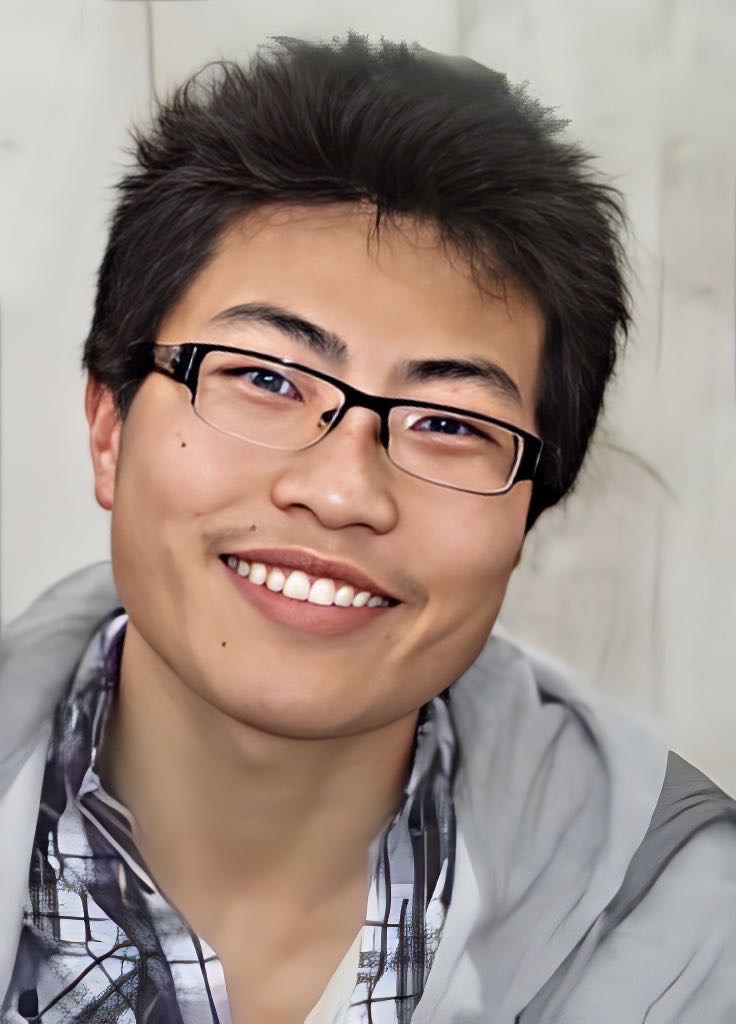}}]{Junfu Pu} received his B.E. degree in Electronic Information Engineering and his Ph.D degree in Information and Communication Engineering from the University of Science and Technology of China (USTC) in 2015 and 2020, respectively. He currently serves as a senior researcher at ARC Lab (Applied Research Center), Tencent. His research interests include sign language recognition/translation/generation, multimedia understanding, multimodal LLMs for image/video applications, vision-language pretraining and search.
\end{IEEEbiography}

\begin{IEEEbiography}
[{\includegraphics[width=1in,height=1.25in,clip,keepaspectratio]{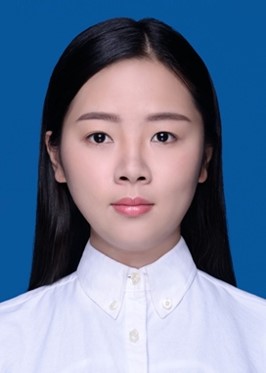}}]{Ying Cheng} received the Ph.D. degree in computer application technology from Fudan University Shanghai, China, in 2023. Her research interests include multimodal analysis, self-supervised learning, and knowledge integration.
\end{IEEEbiography}

\begin{IEEEbiography}
[{\includegraphics[width=1in,height=1.25in,clip,keepaspectratio]{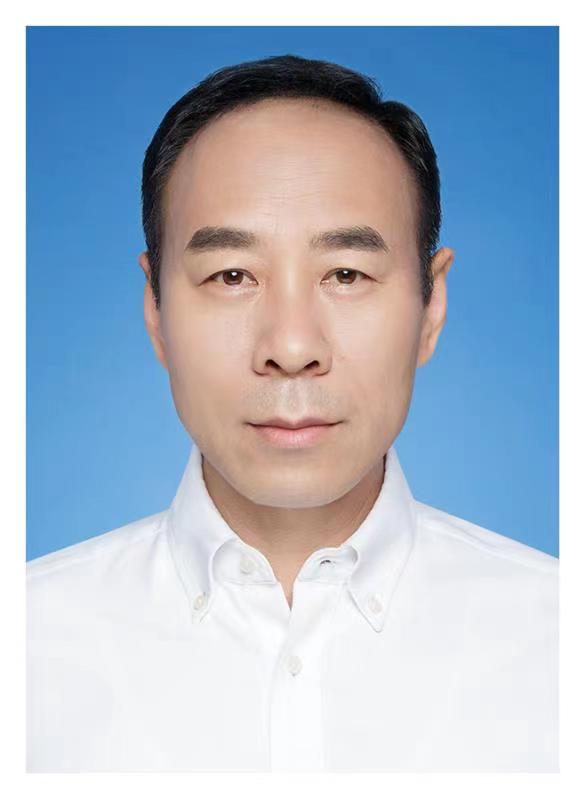}}]{Rui Feng} received the B.S. degree in Industrial Automatic from Harbin Engineering University, Haerbin, China, in 1994, the M.S. degree in Industrial Automatic from Northeastern University, Shenyang, China, in 1997, and the Ph.D. degree in Control Theory and Engineering from Shanghai Jiaotong University, Shanghai, China, in 2003. In 2003, He joined Department of Computer Science and Engineering (now School of Computer Science), Fudan University as an Assistant Professor, and then become Associate Professor and Full Professor. His research interests include medical image analysis, intelligent video analysis, and machine learning.
\end{IEEEbiography}

\begin{IEEEbiography}
[{\includegraphics[width=1in,height=1.25in,clip,keepaspectratio]{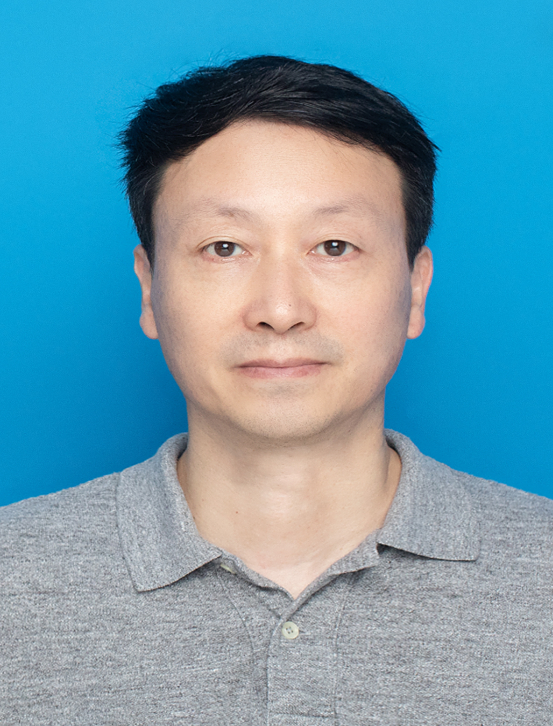}}]{Ying Shan} is a Distinguished Scientist at Tencent, the Director of the ARC Lab at Tencent PCG, and the Director of the Visual Computing Center at Tencent AI Lab. Before joining Tencent, he worked at Microsoft Research as a post-doc researcher, SRI International (Sarnoff Subsidiary) as a Senior MTS, and Microsoft Bing Ads as a Principal Scientist Manager. He has published over 100 papers in top conferences and journals in the areas of computer vision, machine learning, and data mining, served as ACs of CVPR and senior PC of KDD, and holds a number of US/International patents. He is currently leading R\&D efforts in web search, and content AI for a suite of social media and content distribution products.
\end{IEEEbiography}


%\newpage



%\vfill


\end{document}